\documentclass[aps,pra,twocolumn,floats,showpacs]{revtex4}

\usepackage{graphicx}
\usepackage{amsmath}
\usepackage{amssymb}
\usepackage[usenames,dvipsnames]{color}
\usepackage{textcomp}

\usepackage{ulem}

\newcommand{\E}{{\cal E}}

\newcommand{\D}{{\Delta}}

\newcommand{\beq}{\begin{equation}}
\newcommand{\eeq}{\end{equation}}
\newcommand{\bea}{\begin{eqnarray}}
\newcommand{\eea}{\end{eqnarray}}

\begin{document}
\title{Rotational excitation of molecules with long sequences of intense femtosecond pulses}

\date{\today}
\author{M.~Bitter and V.~Milner}
\affiliation{Department of  Physics \& Astronomy and The
Laboratory for Advanced Spectroscopy and Imaging Research
(LASIR), The University of British Columbia, Vancouver, Canada \\}

\begin{abstract}{
We investigate the prospects of creating broad rotational wave packets by means of molecular interaction with long sequences of intense femtosecond pulses. Using state-resolved rotational Raman spectroscopy of oxygen, subject to a sequence of more than 20 laser pulses with peak intensities exceeding $10^{13}$ W/cm$^{2}$ per pulse, we show that the centrifugal distortion is the main obstacle on the way to reaching high rotational states. We demonstrate that the timing of the pulses can be optimized to partially mitigate the centrifugal limit. The cumulative effect of a long pulse sequence results in high degree of rotational coherence, which is shown to cause an efficient spectral broadening of probe light via cascaded Raman transitions.}
\end{abstract}


\maketitle

\section{Introduction}

Control of molecular rotation with ultrashort laser pulses is an active research area of experimental and theoretical molecular science. New regimes of molecular dynamics and molecular interactions can be accessed and studied by means of controlled rotational excitation, e.g. by varying the frequency of molecular rotation and controlling its directionality. The ability to create and shape rotational wave packets is instrumental in achieving high degrees of molecular alignment and, hence, in exploring the electronic structure of molecules. For recent reviews of this broad topic, see \cite{Stapelfeldt2003, Ohshima2010, Fleischer2012}.

Significant progress has been made recently in extending the reach of rotational excitation. An ``optical centrifuge'' \cite{Karczmarek1999, Villeneuve2000} has been used to spin molecules up to the extreme frequencies on the order of 10 THz \cite{Korobenko2014a} and to create coherent superpositions of more than 50 rotational quantum states \cite{Milner2015a}. Owing to the adiabatic mechanism of the centrifuge spinning, controlling the phases of the individual states in such ultra-broad wave packets is not a simple task. An alternative route to generating rotational wave packets with well defined relative phases employs an impulsive excitation by an ultrashort laser pulse.

The width of the wave packet created by a femtosecond laser pulse, acting as an instantaneous rotational ``kick'', can be characterized by a dimensionless  kick strength $P = \D \alpha /(4\hbar) \int \E^2(t) dt$, where $\Delta \alpha $ is the polarizability anisotropy of the molecule, $\E$ is the temporal envelope of the pulse and $\hbar$ is the reduced Planck's constant \cite{Fleischer2009}. Being proportional to the intensity of the laser field, $P$ is limited by the ionization threshold of a given molecule. In a typical case (e.g. N$_{2}$ or O$_{2}$) this means that a wave packet of only a few rotational states can be produced by a single kick.

The ionization limit can be avoided by breaking a single laser pulse into a series of $N$ pulses (or a ``pulse train''). If the pulses in the train are separated by the so-called quantum revival time $T_\text{rev}$, the cumulative kick strength $P_{N}$ of the whole train is the sum of the kick strengths of its individual pulses \cite{Averbukh2001}. Hence, broadening the wave packet can be achieved by increasing the number of pulses, while keeping their intensity below the ionization threshold. Multiple schemes of using long pulse trains (i.e. $N\gg 1$) for the controlled rotational excitation of molecules have been proposed theoretically \cite{Averbukh2001, Leibscher2003, Leibscher2004, Sugny2005, York2008}, and implemented experimentally with sequences of up to eight laser pulses \cite{Cryan2009, Zhdanovich2012, Floss2015, Kamalov2015}.

The revival time is inversely proportional to the second derivative of the rotational energy $E(J)$ with respect to the angular momentum $J$ \cite{Leichtle1996, Seideman1999}. In the case of a rigid rotor, $T_\text{rev}$ reduces to $1/{2cB}$ (where $B$ is the molecular rotational constant and $c$ is the speed of light in vacuum) and does not depend on $J$. However, fast molecular rotation leads to a non-negligible centrifugal distortion and a $J$-dependent revival time. As a result, the accumulation of the total kick strength from pulse to pulse is inhibited, which sets an upper limit to the degree of rotational excitation with periodic pulse trains. Instead of increasing linearly with $N$, the molecular angular momentum oscillates between $J=0$ and the centrifugal limit. The oscillatory behavior has been described theoretically as an analogue of Bloch oscillations in solids \cite{Floss2014} and recently demonstrated experimentally in a gas of rotating molecules \cite{Floss2015} exposed to a pulse train of a moderate strength, $N=8, P_{N}\leq 40$.

Here, we investigate the rotational excitation of oxygen by a pulse train of much higher cumulative kick strength, $N=20, P_{N}\leq 140$. A rigid O$_{2}$ molecule would have been excited to the quantum states with $J\approx 140$, higher than that produced by an optical centrifuge ($J\approx 120$ \cite{Korobenko2014b}). Yet, owing to the centrifugal limit, our high-$P_{N}$ pulse train fails to populate rotational levels with $J> 17$.  We use state-resolved coherent Raman detection to explicitly demonstrate the effect of centrifugal distortion as the main reason for the inhibited rotational ladder climbing.

From the analysis of the rotational Raman spectrum as a function of the timing of pulses in the train, one gains an insight into the possibility of depositing larger amounts of angular momentum despite the centrifugal limit. To achieve this goal, we exploit fractional quantum revivals \cite{Averbukh1989} and modify the pulse sequence so as to expose the molecules to four rotational kicks per $T_\text{rev}$. Such an optimized train allows us to ``pack'' more pulses within the limited amount of time before the centrifugal distortion suppresses further excitation. By fine tuning the timing around each fractional and full revival, we extend our reach from $J\approx17$ to $J\approx 29$, utilizing more efficiently the cumulative strength of a long train with 28 pulses.

Driving molecules to the highly coherent state in a dense medium makes the latter a strong light modulator (for recent reviews, see \cite{Sokolov2003,Baker2011}). In this work, the effect is expressed through the generation of frequency sidebands via the cascaded rotational Raman scattering \cite{Nazarkin1999}. The molecular phase modulation (MPM) imparted on a probe pulse is maximized when the probe delay coincides with a full revival of the rotational wave packet, as well as if its timing is close to that of a fractional revival \cite{Bartels2001}. In the case of a long pulse train optimized for the efficient rotational excitation, the spectral bandwidth is increasingly broadened from pulse to pulse, as has been theoretically studied in \cite{Palastro2012b} and experimentally demonstrated here.

\section{Setup}
\begin{figure}
\centering
 \includegraphics[width=1.0\columnwidth]{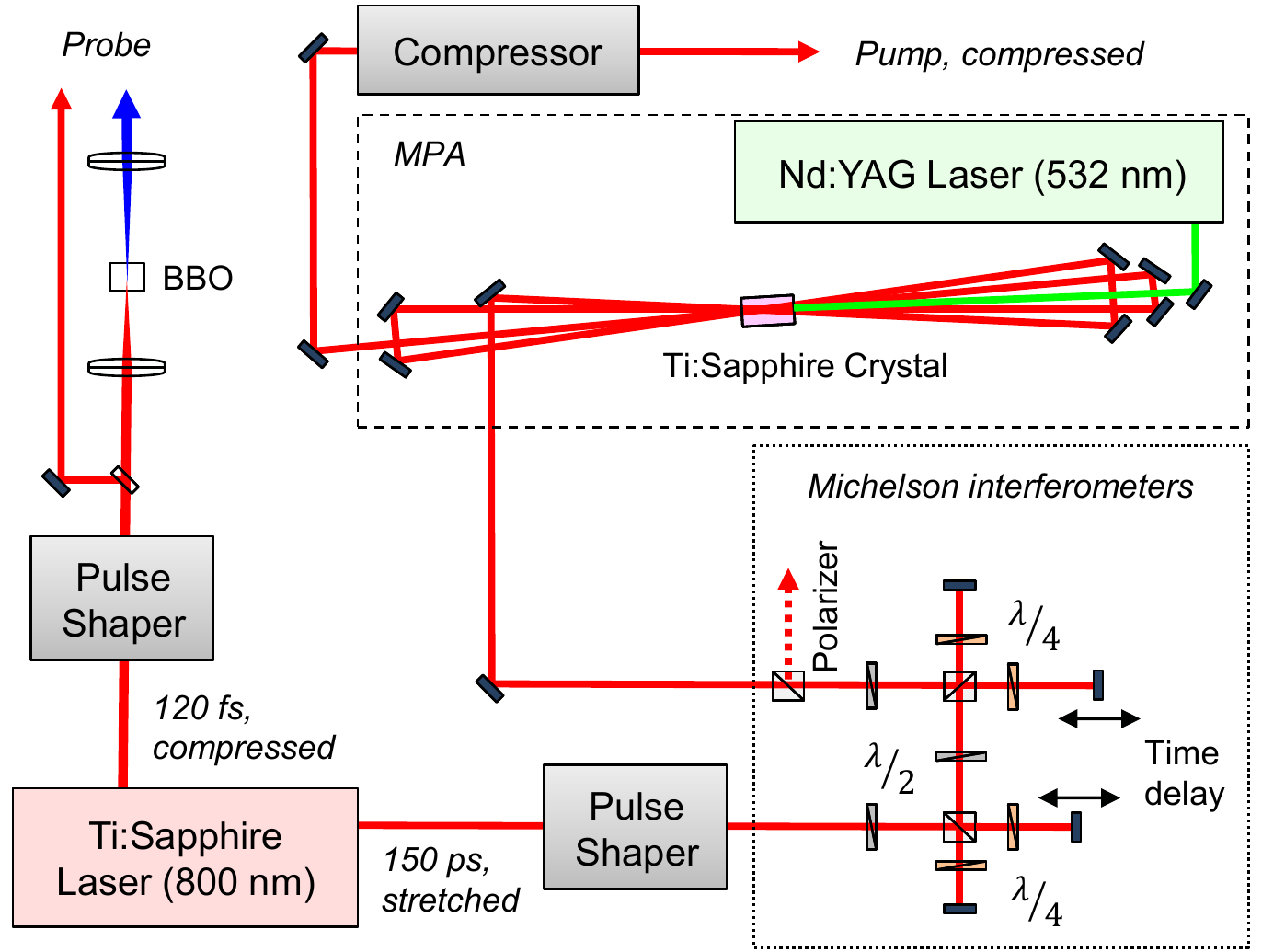}
     \caption{(color online) Experimental setup for the generation of long sequences of amplified femtosecond pulses. See text for details.}
  \vskip -.1truein
  \label{Fig:Setup}
\end{figure}
\begin{figure}
\centering
 \includegraphics[width=1.0\columnwidth]{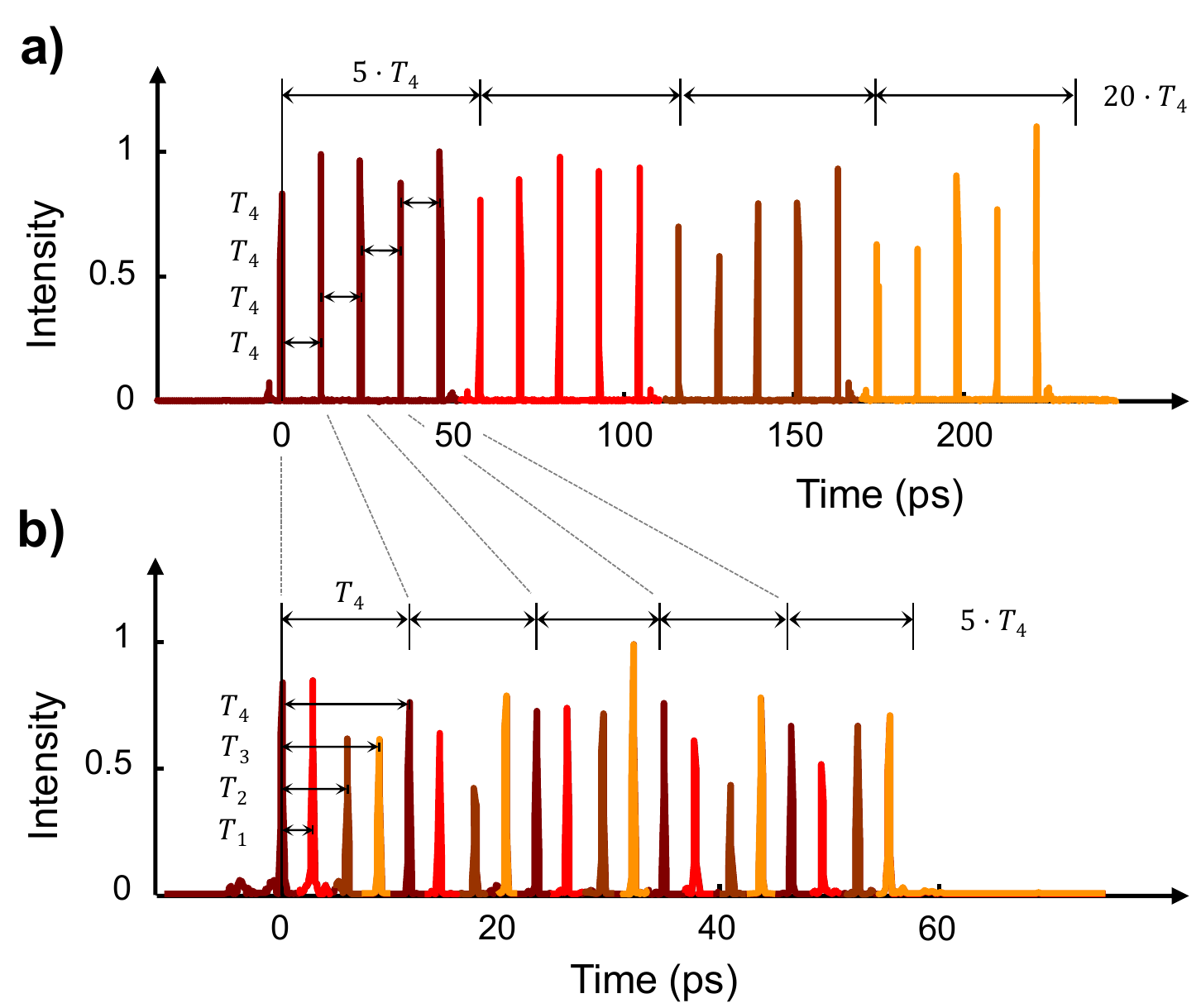}
     \caption{(color online)
     Temporal profile of a \textit{long periodic} pulse train (\textbf{a}) and a \textit{dense aperiodic} pulse train (\textbf{b}), 20 pulses each (note different time scale). The four shades of red correspond to one of the four different pathways through the two nested Michelson interferometers.
     }
  \vskip -.1truein
  \label{Fig:PT}
\end{figure}

Our experimental setup for the generation of long pulse sequences is shown in Fig.~\ref{Fig:Setup} (technical details are discussed elsewhere \cite{Bitter2015}). Briefly, we use a Ti:Sapphire femtosecond laser system (SpitFire Pro, Spectra-Physics) producing pulses of 120~fs full width at half maximum (FWHM) at a central wavelength of 800~nm, 1~KHz repetition rate and 2~mJ per pulse. The pulses are first sent through a pulse shaper, built in a standard `$4f$' geometry \cite{Weiner2000}. Its spectral resolution of 0.04~nm dictates an upper duration limit for the shaped pulse train around 50~ps. The shaper is followed by two Michelson interferometers generating four copies of the shaped train, which are distinguished by four different colors in Fig.~\ref{Fig:PT}.

Producing a long sequence of pulses, whether with a pulse shaper or an interferometer, is typically accompanied by a significant sacrifice of the total energy in the final pulse train. To compensate these energy losses, we amplify the train with a multi-pass amplifier (MPA), as shown in Fig.~\ref{Fig:Setup}. The shaped pulses are passed four times through a Ti:Sapphire crystal, pumped by a Neodymium YAG laser (Powerlite Precision II, Continuum, 800~mJ at 532~nm and 10~Hz). To avoid optical damage and nonlinear propagation effects in air, we shape and amplify chirped pulses before compressing them with a grating compressor in front of the gas cell. The combination of a high resolution pulse shaper, two Michelson interferometers and a MPA allows us to create an arbitrary number of pulses within a time window of about 250~ps, with each pulse reaching energies of 100~$\mu $J. As a result of the nonlinear amplification process, the pulses attain a shorter duration around 100~fs. The standard deviation in the pulse amplitudes within a train is about $20\%$.

We characterize the produced pulse train by cross-correlating it with a single transform-limited 120~fs pulse, which is sent through a separate pulse shaper for dispersion compensation (Fig.~\ref{Fig:Setup}). Two examples of such cross-correlation are shown in Fig.~\ref{Fig:PT}, illustrating the shaping capabilities used in this work. Typically, a periodic pulse train of five pulses with a period $T_4 \approx T_\text{rev}$ is created by the pulse shaper. The interferometers are used to create four copies of this train that can either be added consecutively one after another to form a \textit{long periodic} pulse train of 20 pulses or interleaved with variable timings $T_1$, $T_2$ and $T_3\equiv T_1+T_2$ as shown in Fig.~\ref{Fig:PT}(\textbf{a}) and Fig.~\ref{Fig:PT}(\textbf{b}), respectively. The latter approach enabled us to generate \textit{dense aperiodic} pulse trains optimized for the maximum rotational excitation, as described below.

\begin{figure}
\centering
 \includegraphics[width=1.0\columnwidth]{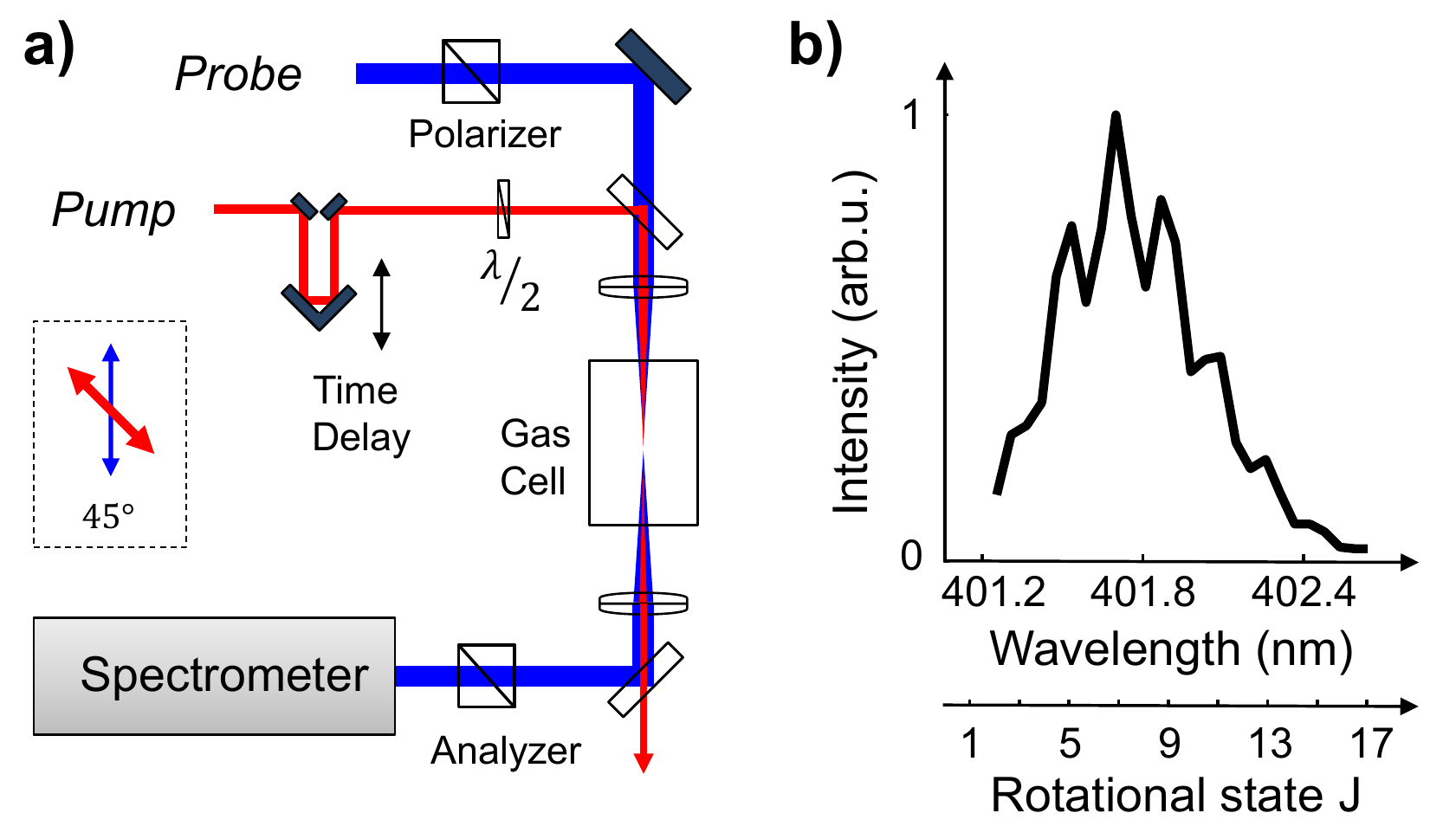}
     \caption{ (color online) (\textbf{a}) Experimental setup for coherent rotational Raman spectroscopy implemented in this work.
     (\textbf{b}) Raman spectrum of room temperature oxygen gas after the rotational excitation with a single weak transform-limited pulse.
     }
  \vskip -.1truein
  \label{Fig:Raman}
\end{figure}

A beta barium borate (BBO) crystal is used to produce probe pulses with the center wavelength of 400.8~nm. Probe pulses are linearly polarized at 45\textdegree\, with respect to the pulses in the train and are combined with the latter on a dichroic beamsplitter. All pulses are focused into a cell containing room temperature oxygen at variable pressure, as shown in Fig.~\ref{Fig:Raman}(\textbf{a}). Molecular alignment, produced by strong pulses in the train, results in the optical birefringence of the gas and a corresponding change in the probe polarization \cite{Renard2003}. The latter is detected by passing the output probe light through an analyzer set at 90\textdegree\, with respect to the initial probe polarization. An optical translation stage is used to scan the probe delay.

We use coherent Raman scattering to detect the rotational coherence, induced by a pulse train in oxygen molecules. When a weak probe pulse follows the train through the cloud of rotating molecules, its spectrum acquires Raman sidebands shifted up or down with respect to the central frequency. The shift of each Raman peak, $\Delta \omega_{J}$, is equal to the frequency spacing between the rotational levels $|J\rangle$ and $|J+2\rangle$, while its magnitude is proportional to the square of the rotational coherence $\rho_{J,J+2}$ \cite{Korobenko2014a}. Using a spectrally narrow probe ($\Delta \lambda \leq 0.15\mathrm{nm}$ FWHM) enables us to resolve individual rotational states of oxygen and to determine the shape of the excited rotational wave packet. Figure~\ref{Fig:Raman}(\textbf{b}) displays such a Raman spectrum taken after the rotational excitation by a single weak kick. In this case, the spectrum reflects the  Boltzmann distribution among the rotational states of molecules. The detected wavelength shift can be translated to the rotational quantum number $J$, shown along the lower horizontal axis. Due to the nuclear spin statistics of oxygen, $J$ can take only odd values. All state-resolved Raman spectra presented in this work were taken with the probe pulse arriving immediately after the last pulse in the train.

\section{Results}

\begin{figure}
\centering
 \includegraphics[width=1.0\columnwidth]{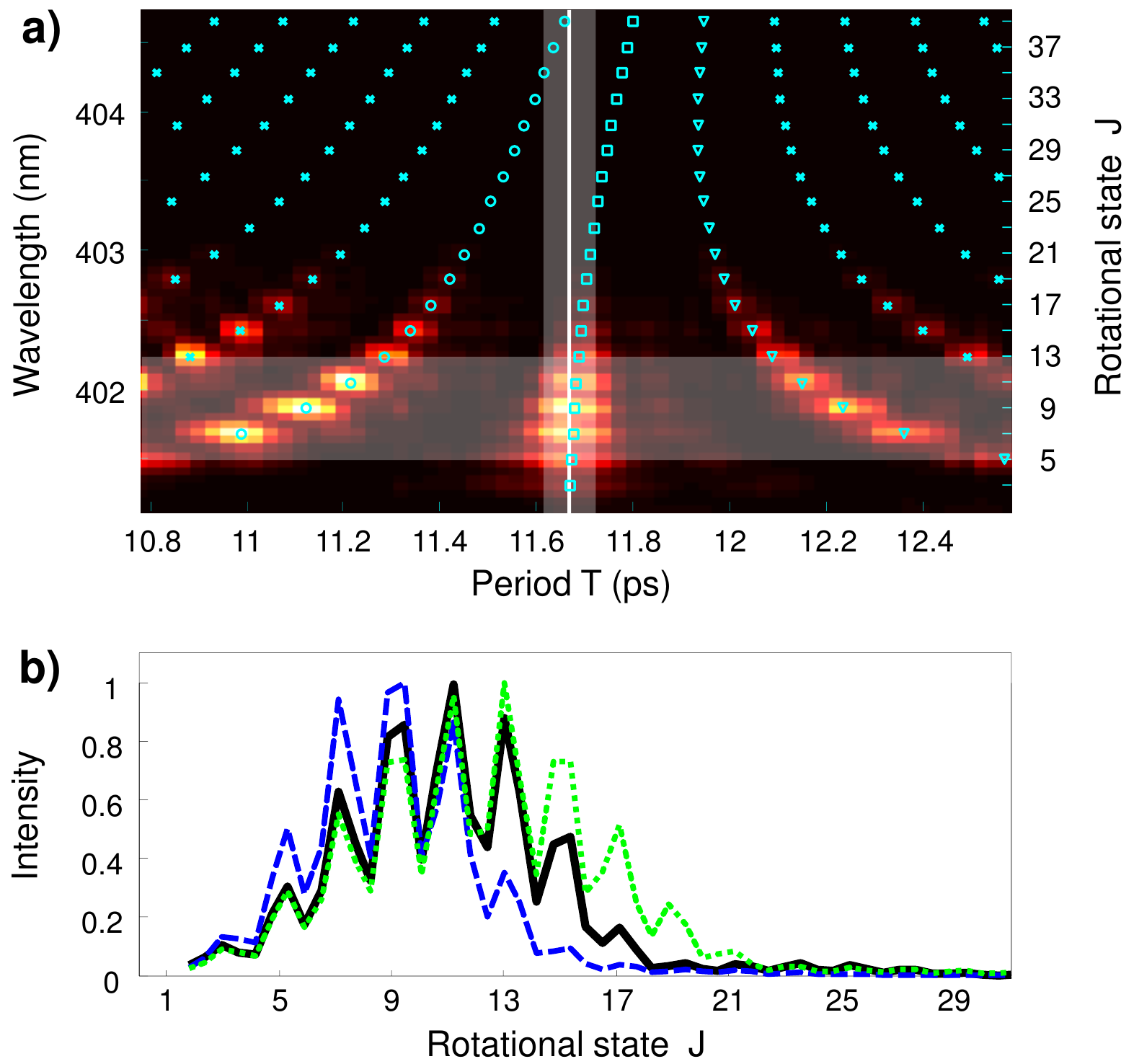}
     \caption{(color online)
     (\textbf{a}) State and time resolved Raman spectrogram of $^{16}$O$_2$ (color coded from dark to bright red) after the excitation by a sequence of five pulses with a period $T$ scanned around the rotational revival time $T_\text{rev}$ (vertical solid line). The vertical shaded band represents the length of our Gaussian pulses (FWHM), whereas the shaded horizonal band covers the rotational quantum numbers (shown on the right vertical axis) corresponding to the thermally populated rotational states (population higher than 10 \%).
     Light blue markers indicate the time moments at which a coherent rotational wave packet consisting of two states, $|J\rangle$ and $|J+2\rangle$, completes an integer multiple of half rotations, $N_J$. The central trajectory (squares) corresponds to $N_J=2J+3$, while the two neighboring sets of circles and triangles represent the cases of $N_J=2J+2$ and $N_J=2J+4$, respectively (see text for details).
     (\textbf{b}) Raman spectra after a periodic sequence of 20 pulses with $T=T_\text{rev}$ - black solid; $T=0.996\cdot T_\text{rev}$ - blue dashed; $T=1.004\cdot T_\text{rev}$ - green dotted.}
  \vskip -.1truein
  \label{Fig:Raman_resonant}
\end{figure}

First, we investigate the rotational coherences created by a \textit{periodic} pulse sequence in a thermal ensemble of oxygen molecules. Figure~\ref{Fig:Raman_resonant}(\textbf{a}) is a 2D plot showing the observed Raman peaks (color coded from dark to bright red) after a train of five relatively weak pulses ($2\cdot 10^{12}\mathrm{W/cm^2}$ or $P=0.5$) as a function of the train period. The apparent pattern of peaks can be interpreted using the following arguments, based on the perturbative regime of light-molecule interaction. The first laser kick in the train induces multiple Raman transitions between the rotational states $|J\rangle$ of a molecule, creating a coherent rotational wave packet $\Psi(t) = \sum_J c_J \ e^{-i E_J t/\hbar} \  |J\rangle $, with the complex amplitudes $c_J$ and rotational energies $E_J$. Consider a wave packet of only two states, $|J\rangle$ and $|J+2\rangle$. It can be assigned a classical rotation period $\tau_J = 2h\left( E_{J+2} - E_J \right)^{-1}$ with Planck's constant $h$. Given the symmetry of an oxygen molecule, the wave function of such a coherent superposition repeats itself every integer multiple of half-rotations, i.e. at times $T_J=N_J\tau_J/2$ ($N_J \in \mathbb{N}$) indicated by light blue markers in Fig.~\ref{Fig:Raman_resonant}(\textbf{a}). If the next laser pulse arrives at this time, it ``kicks'' the molecule in the same direction as the previous pulse, enhancing the rotational excitation of the corresponding $(|J\rangle,|J+2\rangle)$ wave packet.

For a rigid rotor, $E_J=hcBJ(J+1)$ and the rotational period becomes $\tau_J = T_\text{rev}(J+3/2)^{-1}$, with the $J$-independent revival time $T_\text{rev}=1/2cB$. Hence, the number of half-rotations of an excited wave packet during its free evolution between two consecutive laser kicks is $N_J=T/T_\text{rev}\times(2J+3)$.
At the timing known as the ``quantum resonance'', when $T=T_\text{rev}$, 
\textit{all} wave packets perform an integer number of half-rotations with
 $N_J=2J+3$. 
 In Figure~\ref{Fig:Raman_resonant}(\textbf{a}), these times $T_J=(2J+3)\times\tau_J/2$, which are labeled with light blue squares, make up an almost vertical ``trajectory'' originating at the revival time for low values of the rotational quantum number. Thus, a resonant pulse train with the time period $T=T_\text{rev}$ is equally efficient in exciting most molecules in the ensemble, regardless of their angular momentum \cite{Zhdanovich2012}. This is reflected by the highest number of Raman peaks around $T_\text{rev}$ (11.67~ps for $^{16}$O$_2$) marked by the vertical solid line in Fig.~\ref{Fig:Raman_resonant}(\textbf{a}).

Away from the quantum resonance, each rotational wave packet completes an integer number of half-rotations at different time moments. For instance, increasing (decreasing) $N_J$ to the next integer value $2J+4$ ($2J+2$) results in a curved trajectory marked with triangles (circles), along which no two rotational wave packets corresponding to two different values of $J$ are in phase simultaneously. The agreement between the calculated times $T_J$ and the peaks in the measured Raman spectra supports the provided interpretation of the coherent accumulation in the rotational multi-pulse excitation.

Even when the conditions of the quantum resonance are met for the lower rotational states, they no longer hold at higher $J$'s owing to the centrifugal distortion of the molecular bond. With increasing $J$, the centrifugal term in the rotational energy $E_J=hc[BJ(J+1) - DJ^2(J+1)^2]$ (with $D$ being the centrifugal constant) becomes non-negligible, making the resonance $J$-dependent and, therefore, impossible to satisfy for all quantum states simultaneously. In Figure~\ref{Fig:Raman_resonant}(\textbf{a}), this effect is seen through the apparent curving of the resonant trajectory (light blue squares) away from the vertical line at $T=T_\text{rev}$ above $J\approx15$. As a result, the efficiency of the accumulative rotational excitation by a resonant pulse train ($T=T_\text{rev}$) deteriorates with growing $J$. To demonstrate the described centrifugal limit we increased the number of kicks as well as their strength. Fig.~\ref{Fig:Raman_resonant}(\textbf{b}) shows the detected Raman signal generated by the periodic resonant train of 20 strong pulses with pulse intensity of $3\cdot 10^{13}\mathrm{W/cm^2}$ per pulse (black solid line). Despite the increased cumulative kick strength of $P_{20}\approx 140$, the highest excited level remains significantly below that value, $J\approx17\ll P_{20}$. In fact, even though the perturbative analysis used earlier is not applicable in the case of strong pulses, the measured limit agrees well with our conclusions drawn from Figure~\ref{Fig:Raman_resonant}(\textbf{a}), as well as with recent numerical calculations \cite{Floss2014} and experimental observations \cite{Floss2015}.

Utilizing the perturbative picture further, one arrives at a simple method of extending the reach of rotational excitation by a periodic train of femtosecond pulses. As seen in Fig.~\ref{Fig:Raman_resonant}(\textbf{a}), by shifting the train period above the quantum resonance, its overlap with the time of complete half-rotations $T_J$ of the rotational wave packets with higher $J$'s can be improved. However, the degree of such control is rather limited, because detuning the pulse train far from the resonance reduces the excitation efficiency of the initially populated low-$J$ states. The limitation can be analyzed by inspecting the two shaded bands in Fig.~\ref{Fig:Raman_resonant}(\textbf{a}): the vertical one represents the temporal width of our pulses (FWHM), whereas the horizontal one covers the thermally populated rotational states. The length of a continuous set of blue markers under the vertical band tells us about the width of the created rotational wave packet. On the other hand, how many of them are covered by the intersection of the two shaded areas is a qualitative indicator of the number of molecules in the wave packet. As the train period shifts to the right of the quantum resonance, the former factor increases at the expense of the latter.

For the length and strength of our experimental pulse trains, we found the detuning of $+0.4\%$ to result in the highest enhancement of the wave packet's width, while not causing a significant loss of the overall excitation efficiency. As illustrated in Fig.~\ref{Fig:Raman_resonant}(\textbf{b}), by increasing the train period $0.4\%$ above the quantum resonance (green dotted line), we indeed shift the rotational excitation to higher $J$ states. Similarly, setting the train period $0.4\%$ below the resonance (blue dashed line) results in the narrower excited wave packet with the lower ``center of mass''. The idea of small detunings from the resonance was used in \cite{Floss2015} for the detection of Bloch oscillations. By tuning below the resonance and bringing the centrifugal limit to lower $J$ values, one shortens the period of Bloch oscillations, enabling their observation with smaller number of pulses. Bloch oscillations are not visible in this work due to the spatial averaging over the pump beam intensity profile.

We note, that due to the power broadening of the individual rotational transitions, higher pulse intensities should allow farther detunings from the quantum resonance and correspondingly broader rotational wave packets. Stronger pulses effectively push the centrifugal limit higher. In this regime of interaction (which is beyond the scope of this work), the finite width of the laser pulses sets the ultimate limit for the reach of the rotational excitation. Once a molecule rotates too fast for the pulses to act as instantaneous kicks, it does not climb the rotational ladder any further.
\begin{figure}
\centering
 \includegraphics[width=1.0\columnwidth]{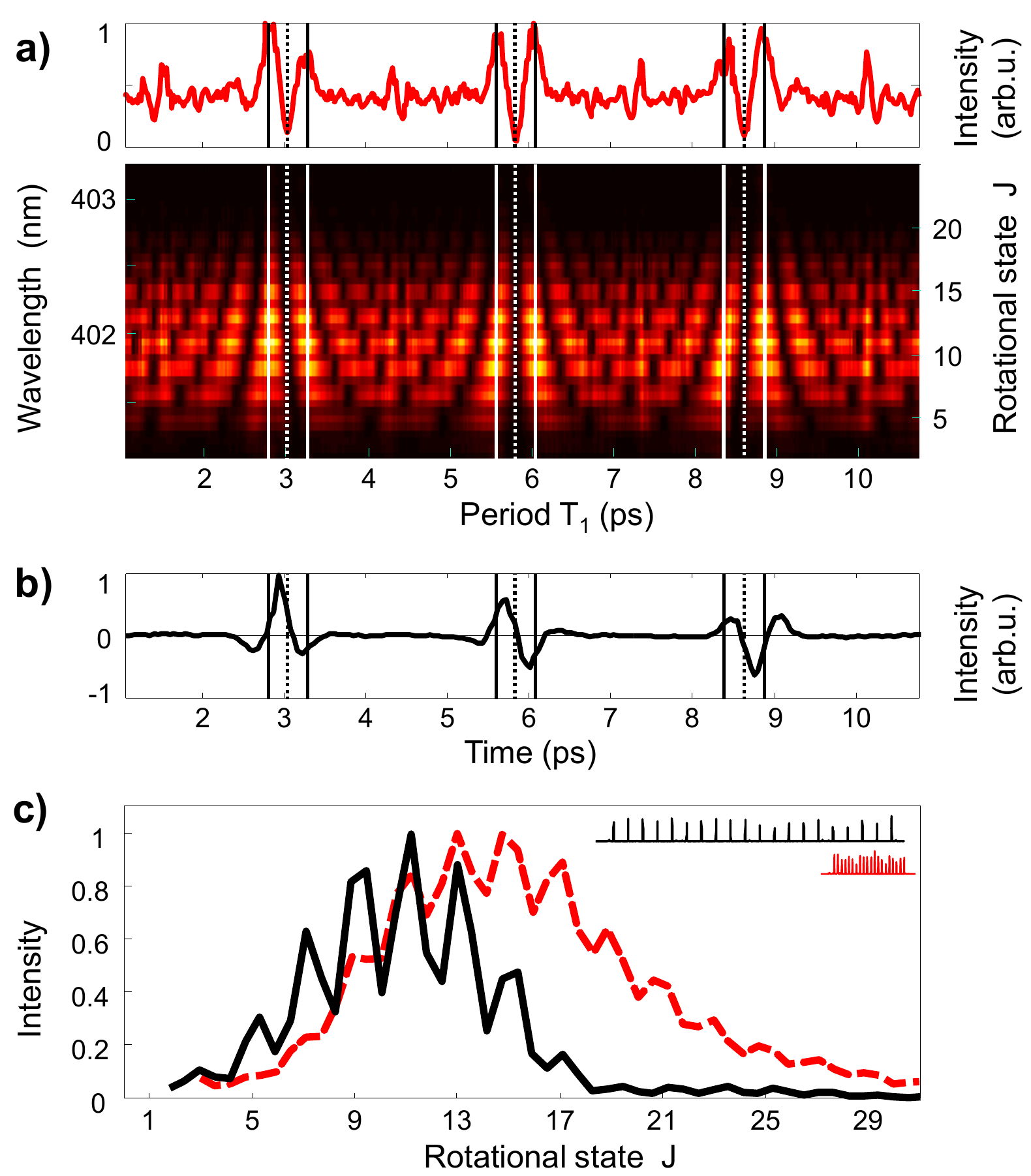}
     \caption{(color online)
         (\textbf{a}) Raman spectrogram for the rotational excitation by a sequence of two identical periodic pulse trains, five pulses each. The period of each train is fixed at $T_\text{rev}$, while the time delay $T_1$ between the two trains is scanned. The integrated signal above reveals the times of maximum total coherence.
         (\textbf{b}) Alignment-induced birefringence signal as a function of the probe delay after a weak transform-limited pump pulse.
     (\textbf{c}) Raman response after the rotational excitation by a sequence of 20 pulses: a periodic pulse train with $T=T_\text{rev}$ (black solid) and an optimized pulse train (red dashed). See text for the optimization procedure. The two trains are illustrated schematically in the upper right corner.}
  \vskip -.1truein
  \label{Fig:Raman_optimized}
\end{figure}

The effect of the centrifugal distortion grows in time as the wave packet spreads out. The dispersion of the revival times with $J$ accumulates, making every next kick less and less efficient \cite{Cryan2009}. Hence, to fully benefit from the large number of pulses, one needs to apply them on as short of a time scale as possible. Here, we utilize fractional revivals \cite{Averbukh1989} and create a tunable non-periodic pulse sequence in order to increase the efficiency of rotational excitation.
\begin{figure*}
\centering
 \includegraphics[width=1\textwidth]{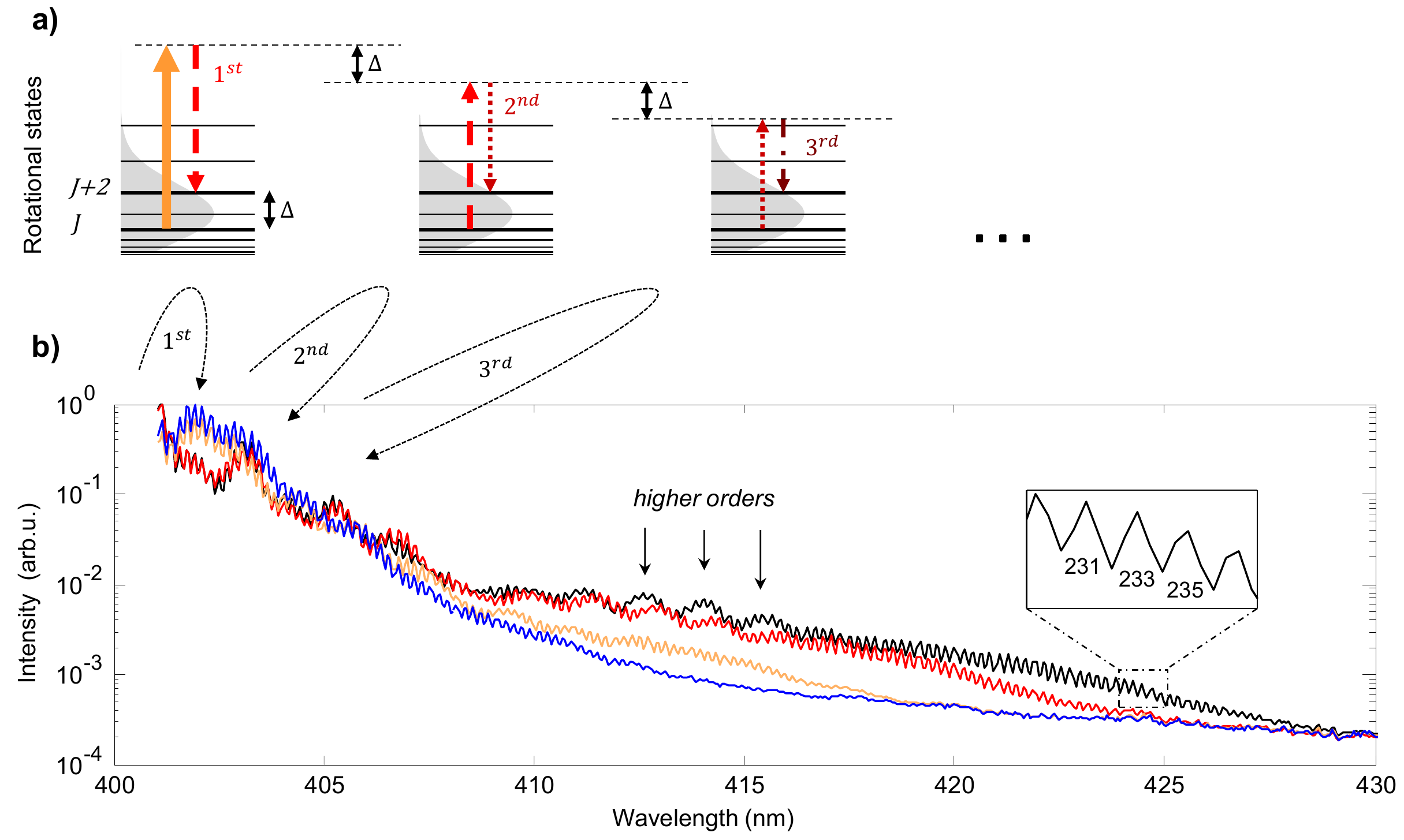}
     \caption{(color online)(\textbf{a}) Simplified illustration of the first three scattering orders in the process of cascaded Raman scattering. Grey profile represents the Boltzmann distribution of rotational population (see text for details).
     (\textbf{b}) Rotational Raman spectra of oxygen gas excited by the optimized train of 28~pulses (see text for the optimization procedure). The four curves correspond to the gas pressures of 1.7~atm (lower blue), 2.4~atm (middle orange), 5.1~atm (higher red) and 6.5~atm (top black). First, second and higher orders of the cascaded Raman scattering are indicated with arrows. More than 235 Raman peaks are observed at higher pressure values, as shown in the inset.}
  \vskip -.1truein
  \label{Fig:CascadedRaman}
\end{figure*}

To start, we demonstrate that utilizing two pulses per every revival time can indeed enhance the excitation, if the timing is chosen appropriately. For that purpose, a train of five pulses with a period $T_4 = T_\text{rev}$ is produced with the pulse shaper. Using a single Michelson interferometer, the train is then split in two parts, which are overlapped in space with the time delay $T_1$ between them (see Fig.~\ref{Fig:PT}(\textbf{b}) for the definition of time intervals $T_1$ and $T_4$). Similarly to the Raman spectrogram in Fig.~\ref{Fig:Raman_resonant}(\textbf{a}), in Fig.~\ref{Fig:Raman_optimized}(\textbf{a}) we plot the observed frequency-resolved Raman signal as a function of the delay $T_{1}$ between the two interleaved periodic pulse trains. As before, strong Raman peaks correspond to the time moments of enhanced rotational excitation. At some of those time moments marked with solid vertical lines, slightly before or after the $(\frac{1}{4}$, $\frac{1}{2}$ and $\frac{3}{4})\times T_\text{rev}$ fractional revivals, the Raman response is simultaneously high for the largest number of rotational states, in exact analogy to the previously discussed response to the periodic excitation at the quantum resonance. We highlight this in the upper panel by plotting the same Raman signal integrated over wavelength. The latter signal is proportional to $\sum_J \left| \rho_{J,J+2} \right| ^2$ and should not be confused with the degree of transient molecular alignment which is plotted in Fig.~\ref{Fig:Raman_optimized}(\textbf{b}) for comparison (the alignment signal was recorded with a 120~fs probe pulse following a single kick).
One can now interpret the optimal $T_1$ values in Fig.~\ref{Fig:Raman_optimized}(\textbf{a}) as the times of the maximum positive derivative of the alignment signal, when the majority of molecules move towards the aligned state. Introducing a second kick at this time results in the enhanced rotational excitation and strong coherent response. In contrast, dashed vertical lines correspond to the maximum negative derivative of the alignment signal, when the majority of molecules move towards the anti-aligned state. A second kick arriving at this time lowers the degree of excitation as reflected by the dips in the integrated coherence signal in panel (\textbf{a}).

The above analysis suggests a possibility of enhancing the rotational excitation by adding up to six pulses per every revival period. Our pulse shaping scheme enables convenient scanning of four pulses per $T_\text{rev}$: near one full and three fractional revivals. We set the four variable delay times to the following optimal values taken from the integrated coherence signal in Fig.\ref{Fig:Raman_optimized}(\textbf{a}): $T_{1}=0.242 \cdot T_\text{rev}$, $T_{2}=0.519 \cdot T_\text{rev}$, $T_{3}=0.761 \cdot T_\text{rev}$ and $T_{4}=1.004\cdot T_\text{rev} = 11.72$ ps, i.e. slightly above the full revival as explained earlier in the text. We note that these optimized delays depend on the kick strength of the pulses in the train. The stronger the pulses, the higher the rotational excitation after each pulse, the larger the centrifugal distortion, the bigger the required shift from every fractional revival.

The result of the excitation by an optimized train is shown in Fig.~\ref{Fig:Raman_optimized}(\textbf{c}). By shortening the duration of the pulse train, while using the same number of pulses ($N=20$, Fig.~\ref{Fig:PT}(\textbf{b})), we extend our reach from J=17 with one pulse per revival (black solid curve) to $J\approx 29$ with four pulse per revival (dashed red curve). Note that although the centrifugal limit of the periodic excitation has been circumvented, the efficiency of the aperiodic optimized pulse train is still well below its accumulated kick strength of $P_N=140$.

At the same time, despite the limited degree of excitation, the induced rotational coherence in the molecular ensemble is rather strong. This can be seen through Raman processes of higher orders, i.e. cascaded Raman scattering, especially pronounced at higher gas pressure. In Fig.~\ref{Fig:CascadedRaman}(\textbf{b}), we plot the spectra of probe pulses following a pulse train of 28 pulses at four different pressures ranging from 1.7 to 6.5~atm. The timing of the pulses in the train has been optimized at 6.5~atm so as to achieve the highest integrated coherence signal at $J>17$, i.e. for maximizing the sum $\sum_{J>17} \left| \rho_{J,J+2} \right| ^2$, by means of the same optimization procedure as described above.

With an optimized pulse sequence, we count up to 250 Raman peaks. Note, that this number does not reflect the rotational quantum number reached by the end of the excitation process (indeed, at such high $J$ values, the separation between the consecutive peaks should have decreased significantly due to the centrifugal distortion). Rather, large frequency shifts are caused by the cascaded Raman scattering giving rise to multiple coherent sidebands \cite{Nazarkin1999}. The process is illustrated schematically in Fig.~\ref{Fig:CascadedRaman}(\textbf{a}). A strong pump pulse (solid orange arrow) creates multiple coherences $\rho_{J,J+2}$ via several two-photon Raman processes (for clarity, only one Raman transition with an average frequency shift $\D$ is shown in the figure). The shape of the resulting Raman spectrum, which consists of a series of $n$ red-shifted peaks, reflects the initial thermal distribution of molecules among the rotational states (grey profile). In the second-order process, all $n$ emitted photons (dashed red arrow) re-scatter off the induced coherences, giving rise to the secondary set of $n^2$ Raman lines (dotted brown arrow), centered around $2\Delta $ from the input laser frequency. The third-order process results in yet another red shift by $\Delta $ (dotted brown to dash-dotted black) and so on. This picture explains the repetitive broad spectral features in Fig.~\ref{Fig:CascadedRaman}(\textbf{b}), e.g. those marked as ``higher orders'', resembling the Raman spectrum observed after a single weak pulse in Fig.~\ref{Fig:Raman}(\textbf{b}). The features are separated by the frequency shift $\D$ corresponding to the peak of the Boltzmann distribution.
The number of Raman processes in each order $m$ of the cascaded scattering grows as $n^m$. We note, that the output spectrum consists of uniformly spaced Raman lines, owing to the repetitive scattering off the same set of coherences, induced by the strongest first-order interaction.

We verified experimentally that the conversion efficiency into the higher-order sidebands increases with the intensity of the pulse train and the number of molecules in the interaction region, both leading to the stronger rotational coherence in the system. Raising the density of molecules by increasing the gas pressure revealed a Bessel-like dependence of the strength of the Raman sidebands on the scattering order number \cite{Nazarkin1999}. For instance, as can be seen in Fig.~\ref{Fig:CascadedRaman}(\textbf{b}), at $P=5.1$ and 6.5~atm, the second-order coherences exceed those induced by the first-order scattering process.

\begin{figure}
\centering
 \includegraphics[width=1.0\columnwidth]{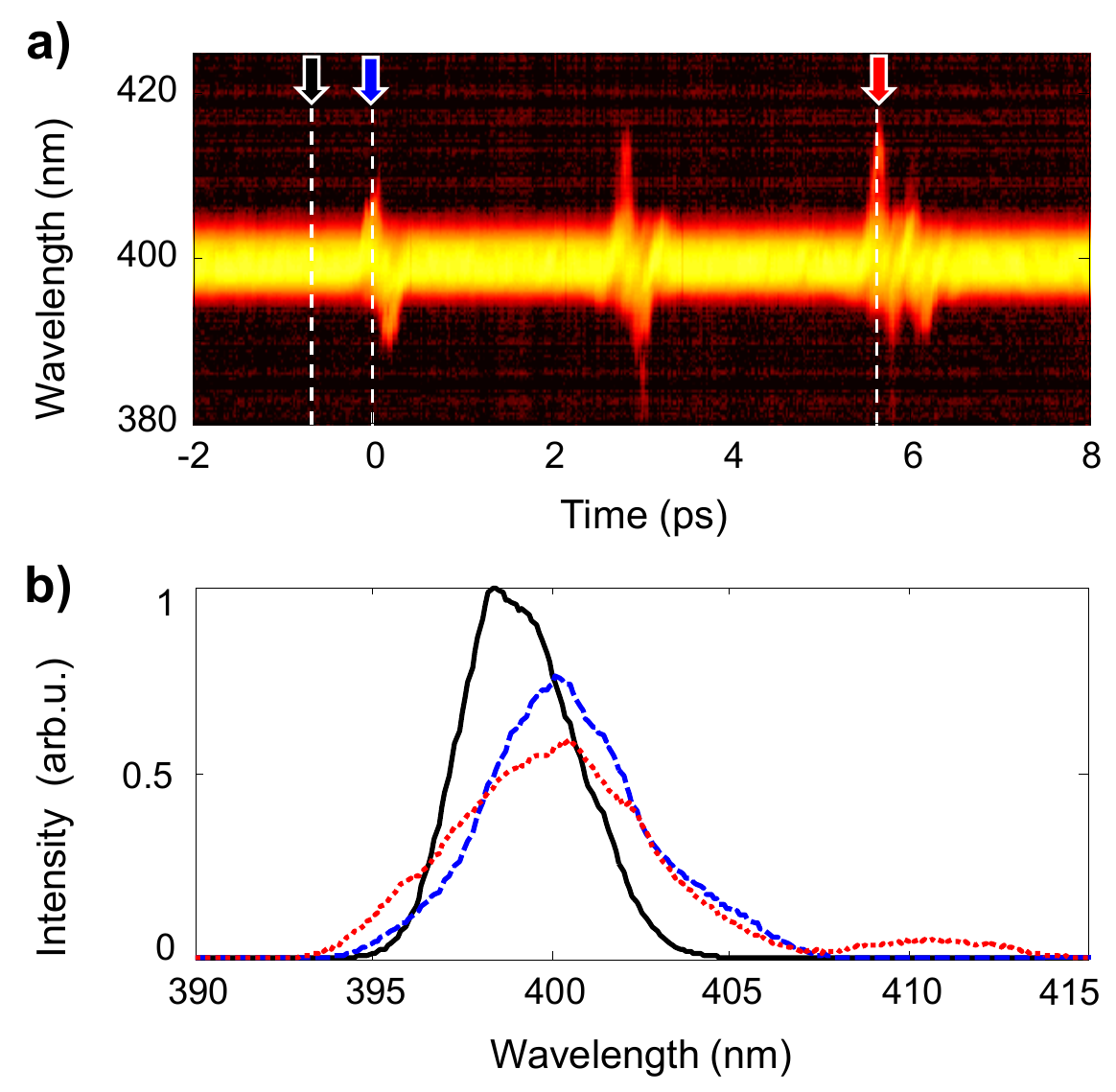}
     \caption{(color online)
     (\textbf{a}) Time-resolved Raman spectrogram of oxygen gas excited by the optimized train of 28~pulses (see text for the optimization procedure) under the pressure of 6.5~atm. (\textbf{b}) Raman spectra at the time moments indicated with arrows in the spectrogram: before the first kick (black solid), at the time of the first kick (blue dashed), and at the time of the third kick (red dotted).}
  \vskip -.1truein
  \label{Fig:Broadening}
\end{figure}

Until now, we have used narrowband probe pulses for the state-resolved detection of the observed spectral broadening driven by the Raman transitions of multiple orders. In the time domain, the process can be described as the transient molecular phase modulation (MPM) owing to the periodically modulated refractive index of the medium. If a femtosecond pulse coincides with a full or a fractional wave packet revival, when the phase modulation is maximized, its frequency bandwidth is broadened \cite{Bartels2001}. The broadening effect has been theoretically predicted to accumulate from pulse to pulse in a periodic sequence of pulses, as long as the train period is equal to the rotational revival time \cite{Palastro2012b}.

To demonstrate this accumulative broadening, we scan the time delay of a 120~fs probe pulse across the optimized pulse train described earlier, while recording the probe spectrum. The results are plotted in Fig.~\ref{Fig:Broadening}(\textbf{a}) as a function of the probe delay with respect to the first pulse in the train. As expected, the spectrum of the probe pulse remains unchanged unless the latter coincides in time with a full or a fractional revival of the wave packet. The further down the pulse train we probe, the broader and more red-shifted the probe spectrum becomes, as illustrated in Fig.~\ref{Fig:Broadening}(\textbf{b}).

\section{Summary}

We have explored the possibilities of efficient rotational excitation of diatomic molecules by long trains of intense femtosecond pulses. In the case of a periodic pulse sequence tuned to the optimal timing of a quantum resonance, the excitation strength is shown to be severely limited by the centrifugal distortion. We have demonstrated that the centrifugal limit can be shifted up to slightly higher values of angular momentum by properly choosing the train period.

To further extend the degree of rotational excitation, we have developed an optimization procedure which produced non-periodic pulse sequences capable of creating broad rotational wave packets surpassing the centrifugal limit. The optimization has been demonstrated with a train of twenty-eight strong femtosecond pulses, with every pulse precisely timed around the corresponding fractional revival of oxygen. Even though the rotational levels, accessed with this technique, are still lower than those expected in the absence of the centrifugal distortion, the induced rotational coherence is enhanced significantly in comparison to the single-pulse rotational excitation. We have demonstrated this enhancement by observing the spectral broadening of probe light due to the strong molecular phase modulation or, equivalently, high-order Raman scattering. The developed method may prove useful for creating ensembles of highly aligned molecules under ambient conditions.

The authors would like to thank Jonathan Morrison for his help with the multi-pass amplifier. This research has been supported by the grants from CFI, BCKDF and NSERC.


\begin{thebibliography}{30}
\expandafter\ifx\csname natexlab\endcsname\relax\def\natexlab#1{#1}\fi
\expandafter\ifx\csname bibnamefont\endcsname\relax
  \def\bibnamefont#1{#1}\fi
\expandafter\ifx\csname bibfnamefont\endcsname\relax
  \def\bibfnamefont#1{#1}\fi
\expandafter\ifx\csname citenamefont\endcsname\relax
  \def\citenamefont#1{#1}\fi
\expandafter\ifx\csname url\endcsname\relax
  \def\url#1{\texttt{#1}}\fi
\expandafter\ifx\csname urlprefix\endcsname\relax\def\urlprefix{URL }\fi
\providecommand{\bibinfo}[2]{#2}
\providecommand{\eprint}[2][]{\url{#2}}

\bibitem[{\citenamefont{Stapelfeldt and Seideman}(2003)}]{Stapelfeldt2003}
\bibinfo{author}{\bibfnamefont{H.}~\bibnamefont{Stapelfeldt}} \bibnamefont{and}
  \bibinfo{author}{\bibfnamefont{T.}~\bibnamefont{Seideman}},
  \bibinfo{journal}{Rev. Mod. Phys.} \textbf{\bibinfo{volume}{75}},
  \bibinfo{pages}{543} (\bibinfo{year}{2003}).

\bibitem[{\citenamefont{Ohshima and Hasegawa}(2010)}]{Ohshima2010}
\bibinfo{author}{\bibfnamefont{Y.}~\bibnamefont{Ohshima}} \bibnamefont{and}
  \bibinfo{author}{\bibfnamefont{H.}~\bibnamefont{Hasegawa}},
  \bibinfo{journal}{Int. Rev. Phys. Chem.} \textbf{\bibinfo{volume}{29}},
  \bibinfo{pages}{619} (\bibinfo{year}{2010}).

\bibitem[{\citenamefont{Fleischer et~al.}(2012)\citenamefont{Fleischer,
  Khodorkovsky, Gershnabel, Prior, and Averbukh}}]{Fleischer2012}
\bibinfo{author}{\bibfnamefont{S.}~\bibnamefont{Fleischer}},
  \bibinfo{author}{\bibfnamefont{Y.}~\bibnamefont{Khodorkovsky}},
  \bibinfo{author}{\bibfnamefont{E.}~\bibnamefont{Gershnabel}},
  \bibinfo{author}{\bibfnamefont{Y.}~\bibnamefont{Prior}}, \bibnamefont{and}
  \bibinfo{author}{\bibfnamefont{I.~Sh.} \bibnamefont{Averbukh}},
  \bibinfo{journal}{Isr. J. Chem.} \textbf{\bibinfo{volume}{52}},
  \bibinfo{pages}{414} (\bibinfo{year}{2012}).

\bibitem[{\citenamefont{Karczmarek et~al.}(1999)\citenamefont{Karczmarek,
  Wright, Corkum, and Ivanov}}]{Karczmarek1999}
\bibinfo{author}{\bibfnamefont{J.}~\bibnamefont{Karczmarek}},
  \bibinfo{author}{\bibfnamefont{J.}~\bibnamefont{Wright}},
  \bibinfo{author}{\bibfnamefont{P.}~\bibnamefont{Corkum}}, \bibnamefont{and}
  \bibinfo{author}{\bibfnamefont{M.}~\bibnamefont{Ivanov}},
  \bibinfo{journal}{Phys. Rev. Lett.} \textbf{\bibinfo{volume}{82}},
  \bibinfo{pages}{3420} (\bibinfo{year}{1999}).

\bibitem[{\citenamefont{Villeneuve et~al.}(2000)\citenamefont{Villeneuve,
  Aseyev, Dietrich, Spanner, Ivanov, and Corkum}}]{Villeneuve2000}
\bibinfo{author}{\bibfnamefont{D.~M.} \bibnamefont{Villeneuve}},
  \bibinfo{author}{\bibfnamefont{S.~A.} \bibnamefont{Aseyev}},
  \bibinfo{author}{\bibfnamefont{P.}~\bibnamefont{Dietrich}},
  \bibinfo{author}{\bibfnamefont{M.}~\bibnamefont{Spanner}},
  \bibinfo{author}{\bibfnamefont{M.~Y.} \bibnamefont{Ivanov}},
  \bibnamefont{and} \bibinfo{author}{\bibfnamefont{P.~B.}
  \bibnamefont{Corkum}}, \bibinfo{journal}{Phys. Rev. Lett.}
  \textbf{\bibinfo{volume}{85}}, \bibinfo{pages}{542} (\bibinfo{year}{2000}).

\bibitem[{\citenamefont{Korobenko
  et~al.}(2014{\natexlab{a}})\citenamefont{Korobenko, Milner, and
  Milner}}]{Korobenko2014a}
\bibinfo{author}{\bibfnamefont{A.}~\bibnamefont{Korobenko}},
  \bibinfo{author}{\bibfnamefont{A.~A.} \bibnamefont{Milner}},
  \bibnamefont{and} \bibinfo{author}{\bibfnamefont{V.}~\bibnamefont{Milner}},
  \bibinfo{journal}{Phys. Rev. Lett.} \textbf{\bibinfo{volume}{112}},
  \bibinfo{pages}{113004} (\bibinfo{year}{2014}{\natexlab{a}}).

\bibitem[{\citenamefont{Milner and Hepburn}(2015)}]{Milner2015a}
\bibinfo{author}{\bibfnamefont{V.}~\bibnamefont{Milner}} \bibnamefont{and}
  \bibinfo{author}{\bibfnamefont{J.~W.} \bibnamefont{Hepburn}},
  \bibinfo{journal}{arXiv:1501.02739}  (\bibinfo{year}{2015}).

\bibitem[{\citenamefont{Fleischer et~al.}(2009)\citenamefont{Fleischer,
  Khodorkovsky, Prior, and Averbukh}}]{Fleischer2009}
\bibinfo{author}{\bibfnamefont{S.}~\bibnamefont{Fleischer}},
  \bibinfo{author}{\bibfnamefont{Y.}~\bibnamefont{Khodorkovsky}},
  \bibinfo{author}{\bibfnamefont{Y.}~\bibnamefont{Prior}}, \bibnamefont{and}
  \bibinfo{author}{\bibfnamefont{I.~Sh.} \bibnamefont{Averbukh}},
  \bibinfo{journal}{New J. Phys.} \textbf{\bibinfo{volume}{11}},
  \bibinfo{pages}{105039} (\bibinfo{year}{2009}).

\bibitem[{\citenamefont{Averbukh and Arvieu}(2001)}]{Averbukh2001}
\bibinfo{author}{\bibfnamefont{I.~Sh.} \bibnamefont{Averbukh}} \bibnamefont{and}
  \bibinfo{author}{\bibfnamefont{R.}~\bibnamefont{Arvieu}},
  \bibinfo{journal}{Phys. Rev. Lett.} \textbf{\bibinfo{volume}{87}},
  \bibinfo{pages}{163601} (\bibinfo{year}{2001}).

\bibitem[{\citenamefont{Leibscher et~al.}(2003)\citenamefont{Leibscher,
  Averbukh, and Rabitz}}]{Leibscher2003}
\bibinfo{author}{\bibfnamefont{M.}~\bibnamefont{Leibscher}},
  \bibinfo{author}{\bibfnamefont{I.~Sh.} \bibnamefont{Averbukh}},
  \bibnamefont{and} \bibinfo{author}{\bibfnamefont{H.}~\bibnamefont{Rabitz}},
  \bibinfo{journal}{Phys. Rev. Lett.} \textbf{\bibinfo{volume}{90}},
  \bibinfo{pages}{213001} (\bibinfo{year}{2003}).

\bibitem[{\citenamefont{Leibscher et~al.}(2004)\citenamefont{Leibscher,
  Averbukh, and Rabitz}}]{Leibscher2004}
\bibinfo{author}{\bibfnamefont{M.}~\bibnamefont{Leibscher}},
  \bibinfo{author}{\bibfnamefont{I.~Sh.} \bibnamefont{Averbukh}},
  \bibnamefont{and} \bibinfo{author}{\bibfnamefont{H.}~\bibnamefont{Rabitz}},
  \bibinfo{journal}{Phys. Rev. A} \textbf{\bibinfo{volume}{69}},
  \bibinfo{pages}{013402} (\bibinfo{year}{2004}).

\bibitem[{\citenamefont{Sugny et~al.}(2005)\citenamefont{Sugny, Keller, Atabek,
  Daems, Dion, Guérin, and Jauslin}}]{Sugny2005}
\bibinfo{author}{\bibfnamefont{D.}~\bibnamefont{Sugny}},
  \bibinfo{author}{\bibfnamefont{A.}~\bibnamefont{Keller}},
  \bibinfo{author}{\bibfnamefont{O.}~\bibnamefont{Atabek}},
  \bibinfo{author}{\bibfnamefont{D.}~\bibnamefont{Daems}},
  \bibinfo{author}{\bibfnamefont{C.~M.} \bibnamefont{Dion}},
  \bibinfo{author}{\bibfnamefont{S.}~\bibnamefont{Guerin}}, \bibnamefont{and}
  \bibinfo{author}{\bibfnamefont{H.~R.} \bibnamefont{Jauslin}},
  \bibinfo{journal}{Phys. Rev. A} \textbf{\bibinfo{volume}{72}},
  \bibinfo{pages}{032704} (\bibinfo{year}{2005}).

\bibitem[{\citenamefont{York and Milchberg}(2008)}]{York2008}
\bibinfo{author}{\bibfnamefont{A.~G.} \bibnamefont{York}} \bibnamefont{and}
  \bibinfo{author}{\bibfnamefont{H.~M.} \bibnamefont{Milchberg}},
  \bibinfo{journal}{Opt. Express} \textbf{\bibinfo{volume}{16}},
  \bibinfo{pages}{10557} (\bibinfo{year}{2008}).

\bibitem[{\citenamefont{Cryan et~al.}(2009)\citenamefont{Cryan, Bucksbaum, and
  Coffee}}]{Cryan2009}
\bibinfo{author}{\bibfnamefont{J.~P.} \bibnamefont{Cryan}},
  \bibinfo{author}{\bibfnamefont{P.~H.} \bibnamefont{Bucksbaum}},
  \bibnamefont{and} \bibinfo{author}{\bibfnamefont{R.~N.}
  \bibnamefont{Coffee}}, \bibinfo{journal}{Phys. Rev. A}
  \textbf{\bibinfo{volume}{80}}, \bibinfo{pages}{063412}
  (\bibinfo{year}{2009}).

\bibitem[{\citenamefont{Zhdanovich et~al.}(2012)\citenamefont{Zhdanovich,
  Bloomquist, Floss, Averbukh, Hepburn, and Milner}}]{Zhdanovich2012}
\bibinfo{author}{\bibfnamefont{S.}~\bibnamefont{Zhdanovich}},
  \bibinfo{author}{\bibfnamefont{C.}~\bibnamefont{Bloomquist}},
  \bibinfo{author}{\bibfnamefont{J.}~\bibnamefont{Flo\ss}},
  \bibinfo{author}{\bibfnamefont{I.~Sh.} \bibnamefont{Averbukh}},
  \bibinfo{author}{\bibfnamefont{J.~W.} \bibnamefont{Hepburn}},
  \bibnamefont{and} \bibinfo{author}{\bibfnamefont{V.}~\bibnamefont{Milner}},
  \bibinfo{journal}{Phys. Rev. Lett.} \textbf{\bibinfo{volume}{109}},
  \bibinfo{pages}{043003} (\bibinfo{year}{2012}).

\bibitem[{\citenamefont{Floss et~al.}(2015)\citenamefont{Floss, Kamalov,
  Averbukh, and Bucksbaum}}]{Floss2015}
\bibinfo{author}{\bibfnamefont{J.}~\bibnamefont{Flo\ss}},
  \bibinfo{author}{\bibfnamefont{A.}~\bibnamefont{Kamalov}},
  \bibinfo{author}{\bibfnamefont{I.~Sh.} \bibnamefont{Averbukh}},
  \bibnamefont{and} \bibinfo{author}{\bibfnamefont{P.~H.}
  \bibnamefont{Bucksbaum}},  \bibinfo{journal}{Phys. Rev. Lett.}
  \textbf{\bibinfo{volume}{115}},
  \bibinfo{pages}{203002} (\bibinfo{year}{2015}).

\bibitem[{\citenamefont{Kamalov et~al.}(2015)\citenamefont{Kamalov, Broege, and
  Bucksbaum}}]{Kamalov2015}
\bibinfo{author}{\bibfnamefont{A.}~\bibnamefont{Kamalov}},
  \bibinfo{author}{\bibfnamefont{D.~W.} \bibnamefont{Broege}},
  \bibnamefont{and} \bibinfo{author}{\bibfnamefont{P.~H.}
  \bibnamefont{Bucksbaum}}, \bibinfo{journal}{Phys. Rev. A}
  \textbf{\bibinfo{volume}{92}}, \bibinfo{pages}{013409}
  (\bibinfo{year}{2015}).

\bibitem[{\citenamefont{Leichtle et~al.}(1996)\citenamefont{Leichtle, Averbukh,
  and Schleich}}]{Leichtle1996}
\bibinfo{author}{\bibfnamefont{C.}~\bibnamefont{Leichtle}},
  \bibinfo{author}{\bibfnamefont{I.~Sh.} \bibnamefont{Averbukh}},
  \bibnamefont{and} \bibinfo{author}{\bibfnamefont{W.~P.}
  \bibnamefont{Schleich}}, \bibinfo{journal}{Phys. Rev. Lett.}
  \textbf{\bibinfo{volume}{77}}, \bibinfo{pages}{3999} (\bibinfo{year}{1996}).

\bibitem[{\citenamefont{Seideman}(1999)}]{Seideman1999}
\bibinfo{author}{\bibfnamefont{T.}~\bibnamefont{Seideman}},
  \bibinfo{journal}{Phys. Rev. Lett.} \textbf{\bibinfo{volume}{83}},
  \bibinfo{pages}{4971} (\bibinfo{year}{1999}).

\bibitem[{\citenamefont{Floss and Averbukh}(2014)}]{Floss2014}
\bibinfo{author}{\bibfnamefont{J.}~\bibnamefont{Flo\ss}} \bibnamefont{and}
  \bibinfo{author}{\bibfnamefont{I.~Sh.} \bibnamefont{Averbukh}},
  \bibinfo{journal}{Phys. Rev. Lett.} \textbf{\bibinfo{volume}{113}},
  \bibinfo{pages}{043002} (\bibinfo{year}{2014}).

\bibitem[{\citenamefont{Korobenko
  et~al.}(2014{\natexlab{b}})\citenamefont{Korobenko, Milner, Hepburn, and
  Milner}}]{Korobenko2014b}
\bibinfo{author}{\bibfnamefont{A.}~\bibnamefont{Korobenko}},
  \bibinfo{author}{\bibfnamefont{A.~A.} \bibnamefont{Milner}},
  \bibinfo{author}{\bibfnamefont{J.~W.} \bibnamefont{Hepburn}},
  \bibnamefont{and} \bibinfo{author}{\bibfnamefont{V.}~\bibnamefont{Milner}},
  \bibinfo{journal}{Phys. Cem. Chem. Phys.} \textbf{\bibinfo{volume}{16}},
  \bibinfo{pages}{4071} (\bibinfo{year}{2014}{\natexlab{b}}).

\bibitem[{\citenamefont{Averbukh and Perelman}(1989)}]{Averbukh1989}
\bibinfo{author}{\bibfnamefont{I.~Sh.} \bibnamefont{Averbukh}} \bibnamefont{and}
  \bibinfo{author}{\bibfnamefont{N.~F.} \bibnamefont{Perelman}},
  \bibinfo{journal}{Phys. Lett. A} \textbf{\bibinfo{volume}{139}},
  \bibinfo{pages}{449} (\bibinfo{year}{1989}).

\bibitem[{\citenamefont{Sokolov and Harris}(2003)}]{Sokolov2003}
\bibinfo{author}{\bibfnamefont{A.~V.} \bibnamefont{Sokolov}} \bibnamefont{and}
  \bibinfo{author}{\bibfnamefont{S.~E.} \bibnamefont{Harris}},
  \bibinfo{journal}{J. Opt. B} \textbf{\bibinfo{volume}{5}},
  \bibinfo{pages}{R1} (\bibinfo{year}{2003}).

\bibitem[{\citenamefont{Baker et~al.}(2011)\citenamefont{Baker, Walmsley,
  Tisch, and Marangos}}]{Baker2011}
\bibinfo{author}{\bibfnamefont{S.}~\bibnamefont{Baker}},
  \bibinfo{author}{\bibfnamefont{I.~A.} \bibnamefont{Walmsley}},
  \bibinfo{author}{\bibfnamefont{J.~W.~G.} \bibnamefont{Tisch}},
  \bibnamefont{and} \bibinfo{author}{\bibfnamefont{J.~P.}
  \bibnamefont{Marangos}}, \bibinfo{journal}{Nat Photon}
  \textbf{\bibinfo{volume}{5}}, \bibinfo{pages}{664} (\bibinfo{year}{2011}).

\bibitem[{\citenamefont{Nazarkin et~al.}(1999)\citenamefont{Nazarkin, Korn,
  Wittmann, and Elsaesser}}]{Nazarkin1999}
\bibinfo{author}{\bibfnamefont{A.}~\bibnamefont{Nazarkin}},
  \bibinfo{author}{\bibfnamefont{G.}~\bibnamefont{Korn}},
  \bibinfo{author}{\bibfnamefont{M.}~\bibnamefont{Wittmann}}, \bibnamefont{and}
  \bibinfo{author}{\bibfnamefont{T.}~\bibnamefont{Elsaesser}},
  \bibinfo{journal}{Phys. Rev. Lett.} \textbf{\bibinfo{volume}{83}},
  \bibinfo{pages}{2560} (\bibinfo{year}{1999}).

\bibitem[{\citenamefont{Bartels et~al.}(2001)\citenamefont{Bartels, Weinacht,
  Wagner, Baertschy, Greene, Murnane, and Kapteyn}}]{Bartels2001}
\bibinfo{author}{\bibfnamefont{R.~A.} \bibnamefont{Bartels}},
  \bibinfo{author}{\bibfnamefont{T.~C.} \bibnamefont{Weinacht}},
  \bibinfo{author}{\bibfnamefont{N.}~\bibnamefont{Wagner}},
  \bibinfo{author}{\bibfnamefont{M.}~\bibnamefont{Baertschy}},
  \bibinfo{author}{\bibfnamefont{C.~H.} \bibnamefont{Greene}},
  \bibinfo{author}{\bibfnamefont{M.~M.} \bibnamefont{Murnane}},
  \bibnamefont{and} \bibinfo{author}{\bibfnamefont{H.~C.}
  \bibnamefont{Kapteyn}}, \bibinfo{journal}{Phys. Rev. Lett.}
  \textbf{\bibinfo{volume}{88}}, \bibinfo{pages}{013903}
  (\bibinfo{year}{2001}).

\bibitem[{\citenamefont{Palastro et~al.}(2012)\citenamefont{Palastro, Antonsen,
  and Milchberg}}]{Palastro2012b}
\bibinfo{author}{\bibfnamefont{J.~P.} \bibnamefont{Palastro}},
  \bibinfo{author}{\bibfnamefont{T.~M.} \bibnamefont{Antonsen}},
  \bibnamefont{and} \bibinfo{author}{\bibfnamefont{H.~M.}
  \bibnamefont{Milchberg}}, \bibinfo{journal}{Phys. Rev. A}
  \textbf{\bibinfo{volume}{86}}, \bibinfo{pages}{033834}
  (\bibinfo{year}{2012}).

\bibitem[{\citenamefont{Bitter and Milner}(2015)}]{Bitter2015}
\bibinfo{author}{\bibfnamefont{M.}~\bibnamefont{Bitter}} \bibnamefont{and}
  \bibinfo{author}{\bibfnamefont{V.}~\bibnamefont{Milner}},
  \bibinfo{journal}{arXiv:1510.07602} (\bibinfo{year}{2015}).  

\bibitem[{\citenamefont{Weiner}(2000)}]{Weiner2000}
\bibinfo{author}{\bibfnamefont{A.~M.} \bibnamefont{Weiner}},
  \bibinfo{journal}{Rev. Sci. Instrum.} \textbf{\bibinfo{volume}{71}},
  \bibinfo{pages}{1929} (\bibinfo{year}{2000}).

\bibitem[{\citenamefont{Renard et~al.}(2003)\citenamefont{Renard, Renard,
  Guérin, Pashayan, Lavorel, Faucher, and Jauslin}}]{Renard2003}
\bibinfo{author}{\bibfnamefont{V.}~\bibnamefont{Renard}},
  \bibinfo{author}{\bibfnamefont{M.}~\bibnamefont{Renard}},
  \bibinfo{author}{\bibfnamefont{S.}~\bibnamefont{Guerin}},
  \bibinfo{author}{\bibfnamefont{Y.~T.} \bibnamefont{Pashayan}},
  \bibinfo{author}{\bibfnamefont{B.}~\bibnamefont{Lavorel}},
  \bibinfo{author}{\bibfnamefont{O.}~\bibnamefont{Faucher}}, \bibnamefont{and}
  \bibinfo{author}{\bibfnamefont{H.~R.} \bibnamefont{Jauslin}},
  \bibinfo{journal}{Phys. Rev. Lett.} \textbf{\bibinfo{volume}{90}},
  \bibinfo{pages}{153601} (\bibinfo{year}{2003}).

\end{thebibliography}

\end{document}